\documentclass[useAMS]{mn2e} 

\usepackage{latexsym,graphicx}

\usepackage{color}

\newcommand\kms{{\rm\,km\,s^{-1}}}

\newcommand\lsun{\rm\,L_\odot}

\def\apgt{\ {\raise-.5ex\hbox{$\buildrel>\over\sim$}}\ }
\def\aplt{\ {\raise-.5ex\hbox{$\buildrel<\over\sim$}}\ }

\title[WR\,72: a born-again planetary nebula]{WR\,72: a born-again planetary nebula with hydrogen-poor knots}
\author[V. V.~Gvaramadze et al.] 
       {V. V.~Gvaramadze,$^{1,2}$\thanks{E-mail: vgvaram@mx.iki.rssi.ru} A. Y.~Kniazev,$^{3,4,1}$ 
       G.~Gr\"afener$^5$ and N.~Langer$^5$ \\
        $^1$Sternberg Astronomical Institute, Lomonosov Moscow State University, Universitetskij Pr. 13, Moscow 119234, Russia\\
        $^2$Space Research Institute, Russian Academy of Sciences, Profsoyuznaya 84/32, 117997 Moscow, Russia \\
        $^3$South African Astronomical Observatory, PO Box 9, 7935 Observatory, Cape Town, South Africa \\
        $^4$Southern African Large Telescope Foundation, PO Box 9, 7935 Observatory, Cape Town, South Africa \\
        $^5$Argelander-Institut f\"ur Astronomie, Auf dem H\"ugel 71, 53121 Bonn, Germany \\    
        }
\begin{document}

\date{Accepted 2019 December 21. Received 2019 December 20; in original form 2019 November 26.}


\maketitle

\label{firstpage}

\begin{abstract}
We report the discovery of a handful of optical hydrogen-poor knots in the central part of an 
extended infrared nebula centred on the [WO1] star WR\,72, obtained by spectroscopic and imaging 
observations with the Southern African Large Telescope (SALT). {\it Wide-field Infrared Survey 
Explorer} ({\it WISE}) images of the nebula show that it is composed of an extended almost 
circular halo (of $\approx6$ arcmin or $\approx2.4$ pc in diameter) and an elongated and 
apparently bipolar inner shell (of a factor of six smaller size),  within which the knots are 
concentrated. Our findings indicate that WR\,72 is a new member of the rare group of hydrogen-poor 
planetary nebulae, which may be explained through a very late thermal pulse of a post-AGB star, 
or by a merger of two white dwarfs.
\end{abstract}


\section{Introduction}
\label{sec:intro}

\begin{figure*}
    \begin{center}
	\includegraphics[width=18cm]{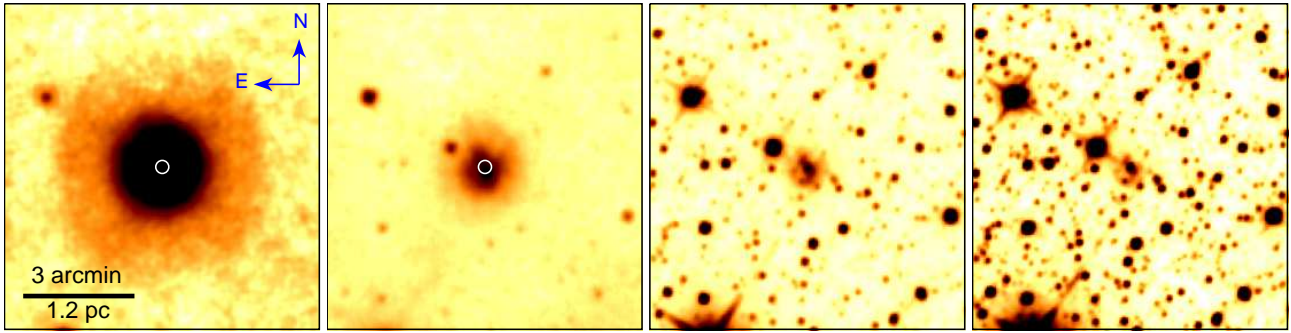}
	\end{center}
	\caption{From left to right: {\it WISE} 22, 12, 4.6 and 3.4\,$\micron$ images of WR\,72 (marked 
	by a circle) and its circumstellar nebula. The orientation and the scale of the images are the same. 
	}
	\label{fig:neb}
\end{figure*}

\begin{figure*}
\centering
    \begin{center}
	\includegraphics[width=18cm]{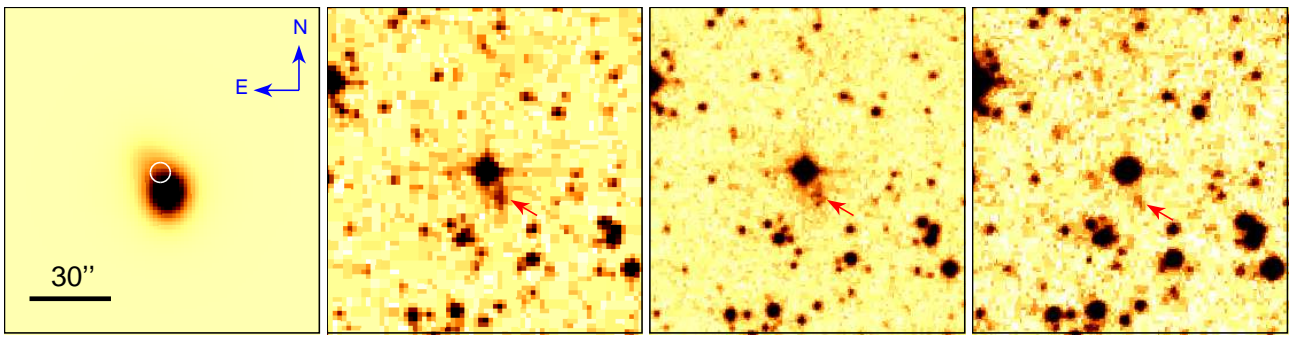}
	\end{center}
	\caption{From left to right: {\it WISE} 22\,$\micron$, DSS-II blue-band, and SSS blue- and red-band 
	images of the close vicinity of WR\,72 (the position of this star in the 22\,$\micron$ image is shown 
	by a circle). The optical knot detected to the soutwest of WR\,72 is indicated by arrow. Note that 
	the peak of the 22\,$\micron$ emission coincides with the optical knot. The orientation and the scale 
	of the images are the same. 
	}
	\label{fig:knot}
\end{figure*}

Among planetary nebulae (PNe) there is a rare group of objects showing hydrogen-poor (H-poor) 
material in the central region of older and larger hydrogen-rich nebula (Jacoby 1979; Hazard
et al. 1980; Jacoby \& Ford 1983; see also Zijlstra 2002 for a review). In the two best-studied 
members of this group, A30 (PN A66 30) and A78 (PN A66 78), the H-poor material appears as a fan 
of knots with cometary tails, stretched radially from the central star (e.g. Borkowski et al. 
1993, 1995; Fang et al. 2014). The 3D distribution of the knots suggests the existence of an 
equatorial disc-like structure and polar outflows with individual knots moving outwards with 
velocities of $\sim100 \, \kms$ in the equatorial ring and at a factor of several higher velocities 
in the polar directions (e.g. Borkowski et al. 1995; Chu et al. 1997; Fang et al. 2014).

The origin of the knots could be understood as the result of instabilities developed during 
the interaction between the current fast wind and ionizing emission of the central star 
of PN (CSPN) with the slow dense H-poor material lost by the star shortly before (e.g. Fang et al. 
2014). Some of the knots have cold, neutral cores, as evidenced by the detection of molecular 
emission at (sub)millimeter wavelengths (Tafoya et al. 2017). The optical emission from these knots 
comes from the ionized skin around the neutral core. The material photoevaporated and ablated from
the knots could be hot and dense enough to produce observable X-ray emission (Chu et al. 1997; 
Guerrero et al. 2012; Toal\'a et al. 2015).

It is believed that the origin of the H-poor material is due to very late thermal pulse (VLTP; 
Fujimoto 1977; Sch\"onberner 1979; Iben et al. 1983), leading to the (almost) total burning of the 
remaining hydrogen shell in the star (e.g. Herwig et al. 1999). Under the influence of the VLTP, the 
CSPN expands and cools, and finds itself again at the asymptotic giant branch (AGB). During this 
``born-again" episode the CSPN develops a slow H-poor wind, which is fragmented in knots when the 
star leaves the AGB for the second time and again starts to produce fast wind.

On the other hand, the axially symmetric distribution of the knots suggests that the origin of the 
H-poor PNe could be somehow related to the binarity of their central stars or to a combination of the 
binary evolution and the VLTP in one of the binary components (e.g. Harrington 1996; Wesson et al. 
2008; de Marco 2008; Lau et al. 2011; Gvaramadze et al. 2019b). Revealing new members of the group of 
H-poor PNe is therefore of high importance for understanding their nature and for the theory of low-mass 
star evolution in general. In this paper, we show that WR\,72 is a new member of this rare group.

\section{WR\,72 and its cirumstellar nebula}
\label{sec:neb}

WR\,72 (or Sand\,3; Sanduleak 1971) is a hydrogen-deficient, ultrahigh-excitation star exhibiting 
a Wolf-Rayet-type spectrum (Barlow, Blades \& Hummer 1980). Originally, it was listed in the Sixth 
Catalogue of Galactic Wolf-Rayet stars (van der Hucht et al. 1981) as a peculiar WC4 star, but was 
omitted from the later editions of the catalogue after it was reclassified as a CSPN by Barlow \&
Hummer (1982). The main reasons for this reclassification are the high-latitude ($b\approx12\degr$) 
location of the star and the presence of the N\,{\sc v} $\lambda$1238\,\AA \, line in its UV spectrum, 
which is not expected in massive stars as they completely destroy nitrogen in the early stages of 
helium burning. According to the classification scheme of Crowther et al. (1998), WR\,72 ia a [WO1] 
star. In the SIMBAD data base\footnote{http://simbad.u-strasbg.fr/simbad/} WR\,72 is indicated as 
a PN.

Since no PN associated with WR\,72 was found at the time, Barlow \& Hummer (1982) suggested that it has 
already dispersed in the interstellar medium and is no longer detectable. Later on, van der Hucht et
al (1985) detected an unresolved infrared (IR) source\footnote{Note that the {\it IRAS} Small Scale 
Structure Catalog (Helou \& Walker 1988) gives for this source an angular diameter of 1.3 arcmin at 
25\,$\micron$.} centred on WR\,72 using 50 and 100\,$\micron$ data from the {\it Infrared Astronomical 
Satellite} ({\it IRAS}) Chopped Photometric Channel instrument, and presented the {\it IRAS}
flux densities for this source at 12, 25, 60 and 100\,$\micron$ (taken from the {\it IRAS} Point Source 
Catalogue). Using these flux densities, they constructed spectral energy distribution of the IR source 
and treated it as a dust shell with a temperature within the range of temperatures found in PNe. 
The subsequent search for an optical nebula 
around WR\,72 using a narrowband H\,$\alpha$ filter gave a negative result (Marston et al. 1994). 
Feibelman (1996) found evidence of nebular emission lines (produced by triply or higher ionized species) 
in the {\it International Ultraviolet Explorer} ({\it IUE}) spectrum of WR\,72, but 
regarded this star as a ``central star without planetary nebula" because no classical PN has ever been 
detected. The nebula also was not detected in the {\it Hubble Space Telescope} ({\it HST}) UV spectrum 
of WR\,72 (Keller et al. 2014), which again was considered as a prove that it has already dispersed. 

Still, the nebula around WR\,72 exists! We found it as a by-product of our search for massive 
evolved stars through the detection of their circumstellar nebulae (Gvaramadze et al. 2010,
2012). The nebula was detected in the data of the all sky survey carried out with the {\it Wide-field Infrared 
Survey Explorer} (WISE; Wright et al. 2010), which provides images at four wavelengths: 22, 12, 4.6 and 
3.4\,$\micron$, with angular resolution of 12.0, 6.5, 6.4 and 6.1 arcsec, respectively, and it is 
clearly seen at all four {\it WISE} wavebands. The literature search revealed that
the nebula was first detected by Griffith et al. (2015), who used the {\it WISE} data to search for 
objects with extreme mid-IR colours, which might be associated with extraterrestrial civilizations 
with large energy supplies. Since the SIMBAD data base indicates a PN at the position of the nebula 
it was graded by Griffith et al. (2015) as a source whose nature is well understood, and correspondingly 
no images of the nebula were presented.
 
In Fig.\,\ref{fig:neb}, we show for the first time the {\it WISE} images of the nebula around WR\,72. 
At 22\,$\micron$ it appears as an almost circular (but slightly elongated in the north-south direction) 
diffuse halo (of angular diameter of $\approx6$ arcmin) surrounding a core of bright emission 
centred around WR\,72. Interestingly, the peak of this emission is shifted to the
southwest of WR\,72 (see Fig.\,\ref{fig:knot}). At 12\,$\micron$ the halo is not visible, while the 
central core is resolved in an asymmetric structure of angular size of $\approx1$ arcmin surrounded 
by diffuse emission of the same size as the bright core in the 22\,$\micron$ image. Within this diffuse 
emission one can distinguish several radial spokes stretched to the northeast and northwest. At shorter 
wavelengths (4.6 and 3.4\,$\micron$) the nebula appears as an elongated (apparently bipolar) shell of 
angular size of $\approx1$ arcmin. The {\it Gaia} second data release (DR2; Gaia Collaboration
et al. 2018) parallax of $0.705\pm0.039$ mas places WR\,72 at the distance of $\approx1.42\pm0.08$ kpc. 
At this distance, the linear sizes of the halo and the central shell are, respectively, $\approx2.4$ 
and 0.4 pc. 

We did not find an optical counterpart to the IR halo in the available optical sky surveys, but noticed 
a knot of diffuse emission at $\approx10$ arcsec to the southwest from WR\,72 in the blue-band image 
(see Fig.\,\ref{fig:knot}) from the Digitized Sky Survey\,II (DSS-II; McLean et al. 2000). The non-stellar 
origin of this emission is more obvious in the blue-band image (see Fig.\,\ref{fig:knot}) from the 
SuperCOSMOS Sky Survey (SSS; Hambly et al. 2001). The gleam of the knot is also visible in the SSS red-band 
image (see Fig.\,\ref{fig:knot}) and can be discerned in the DSS-II red-band image (not show in 
Fig.\,\ref{fig:knot}). Comparison of the IR and optical images shows that the knot coincides with the 
peak of the 22\,$\micron$ emission (see Fig.\,\ref{fig:knot}).

The SIMBAD database lists the {\it IRAS} source 16032$-$3537 at $\approx9$ arcsec to the southwest from 
WR\,72 and notes that ``IRAS\,16032$-$3537 is not WR\,72". The detection of the IR nebula around WR\,72, 
whose brightness peaks at the position of the {\it IRAS} source, implies that this source is in fact 
associated with WR\,72.

\begin{figure*}
\begin{center}
\includegraphics[width=15cm, clip=]{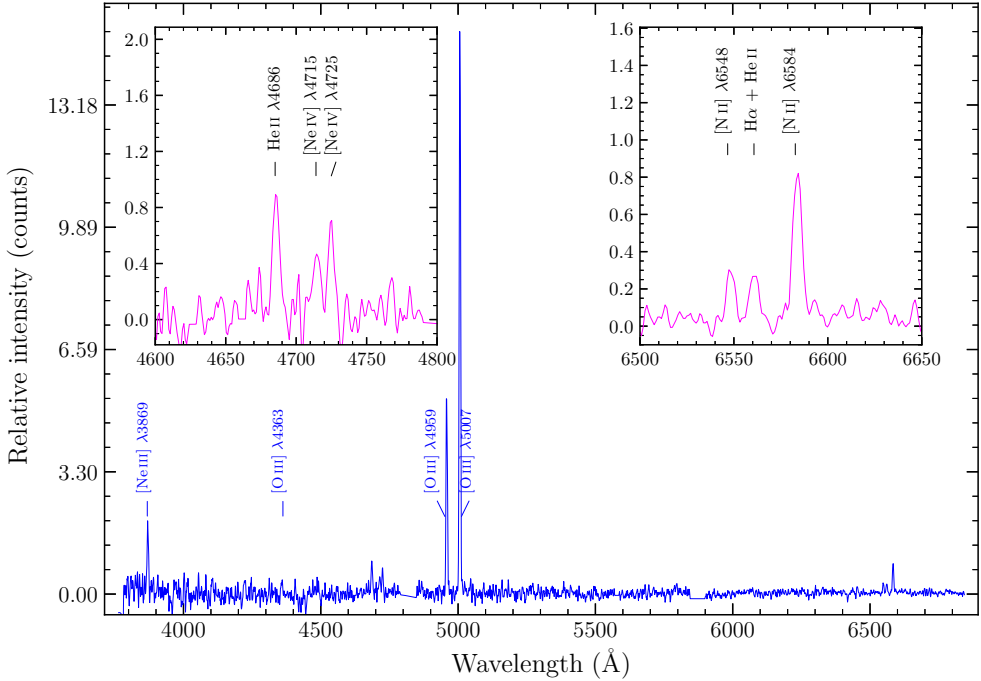}
\end{center}
\caption{1D spectrum of the optical knot. The gaps in the spectrum at $\approx4800$ and 5900\,\AA \, are 
CCD gaps.}
\label{fig:rss}
\end{figure*}

\section{SALT observations}
\label{sec:obs}

To reveal the nature of the optical knot, we observed it with the South African Large Telescope 
(SALT; Buckley et al. 2006; O’Donoghue et al. 2006). First, we took its spectrum using the Robert 
Stobie Spectrograph (RSS; Burg et al. 2003; Kobulnicky et al. 2003) in the long-slit spectroscopy 
mode on 2019 July 16. A 8\,arcmin$\times$1.5 arcsec slit was placed on the knot at the position angle 
PA=45\degr \, (see Fig.\,\ref{fig:salt}). Two 1000s exposures were obtained using the PG900 grating.
This spectral setup covers a wavelength range of $3750-6850$\,\AA \, with a spectral 
resolution FWHM of 5.6$\pm$0.3~\AA. The seeing during these observations 
was $\approx2.1$ arcsec. A spectrum of an Ar comparison arc was obtained immediately after that to 
calibrate the wavelength scale. For the relative flux calibration a spectrophotometric 
standard star EG\,21 (Baldwin \& Stone 1984) was observed with the same spectral set-ups during 
nearest twilights as a part of the SALT calibration plan. The absolute flux calibration is not 
possible with SALT because the unfilled entrance pupil of the telescope moves during the observations.

Then, we observed WR\,72 using the RSS imaging mode on 2019 August 6. A 900s exposure was taken with 
the filter PI05060 ($\lambda_{\rm c}$=5071.5\,\AA, FWHM=110.5\,\AA) to cover the 
[O{\sc iii}]~$\lambda$5007~\AA\ emission line. Also, a 900s exposure was taken with the filter PI05145 
($\lambda_{\rm c}$=5152.1\,\AA, FWHM=109.2\,\AA) to obtain an image free from emission lines, to 
remove stellar confusion. The seeing in these observations was $\approx1.6$ arcsec. Both images were 
binned by a factor of 4 to get the final spatial sampling of 0.5 arcsec\,pixel$^{-1}$.

The primary reduction of the RSS data was done with the SALT science pipeline (Crawford et al. 2010). 
The subsequent long-slit data reduction was carried out in the way described in Kniazev et al. (2008).

\section{Spectrum of the optical knot}
\label{sec:knot}

The 1D spectrum of the optical knot is presented in Fig.\,\ref{fig:rss}. It was extracted by summing 
up, without any weighting, all rows in the brightest part of the knot (i.e. in the angular distance
interval from $-7$ arcsec to $-12.5$ arcsec; see Fig.\,\ref{fig:vel}). The spectrum is dominated by the strong 
[O{\sc iii}]~$\lambda\lambda$4959, 5007~\AA\ doublet. The next strongest emissions are due to the 
[Ne\ {\sc iii}] $\lambda$3869\,\AA, He\ {\sc ii} $\lambda$4686\,\AA \, and [N\,{\sc ii}] $\lambda$6584 
\AA \, lines. The emission line at the position of the He\,{\sc ii} $\lambda$6560 \AA \, + H\,$\alpha$
blend is very weak. 

The detected lines are listed in Table~\ref{tab:int} along with their observed 
intensities (normalized to He\ {\sc ii} $\lambda$4686\,\AA) and the reddening-corrected line intensity 
ratios (derived for $E(B-V)=0.4$ mag; Keller et al. 2014). The electron temperature diagnostic line 
[O\,{\sc iii}]\,$\lambda$4363 \AA \, is not visible in the spectrum. For this line we provide the upper 
limit assuming that its intensity is equal to the $1\sigma$ noise level. 

\begin{table}
\centering{
\caption{Line intensities of the optical knot.}
\label{tab:int}
\begin{tabular}{lcc} 
\hline
$\lambda_{0}$(\AA) Ion                  & $F(\lambda)/F$(He\,{\sc ii} 4686) & $I(\lambda)/I$(He\,{\sc ii} 4686) \\ 
\hline
3868\ [Ne\ {\sc iii}]                  & 2.33$\pm$0.20 & 3.14$\pm$0.28 \\
4363\ [O\ {\sc iii}]                  & $<0.14$      &$<0.16$        \\
4686\ He\ {\sc ii}                    & 1.00$\pm$0.09 & 1.00$\pm$0.09 \\
4715\ [Ne\ {\sc iv]}                  & 0.50$\pm$0.07 & 0.49$\pm$0.07 \\
4725\ [Ne\ {\sc iv]}                  & 0.61$\pm$0.09 & 0.60$\pm$0.09 \\
4959\ [O\ {\sc iii}]                  & 7.00$\pm$0.47 & 6.41$\pm$0.43 \\
5007\ [O\ {\sc iii}]                  &21.02$\pm$1.42 &18.94$\pm$1.29 \\
6548\ [N\ {\sc ii}]                    & 0.22$\pm$0.05 & 0.13$\pm$0.03 \\
6560\ He\ {\sc ii}+H\,$\alpha$        & 0.21$\pm$0.05 & 0.13$\pm$0.03 \\
6584\ [N\ {\sc ii}]                    & 0.94$\pm$0.08 & 0.57$\pm$0.06 \\
\hline
\end{tabular}
  }
\end{table}

The very low intensity of the He\,{\sc ii}+H\,$\alpha$ blend suggests that the knot is H-poor. To constrain 
its H/He abundance ratio, we compare the observed H\,$\alpha$/He\,{\sc ii}\,4686 line ratio with theoretical 
estimates. The observed He\,{\sc ii} emission originates in a region where helium is fully ionized to 
He$^{++}$. In the following we obtain an upper limit for the H abundance by estimating the He\,{\sc ii} and
H\,{\sc i} emission originating from this region alone.

We compute the strengths of the components of the H\,$\alpha$+He\,{\sc ii} blend relative to He\,{\sc ii}\,4686 
using theoretical line emissivities for hydrogenic ions from Martin (1988), and the scaling relation between 
line emissivities $j$ of hydrogenic ions with core charge $Z$ at electron temperature $T$:
\begin{equation}
j_{n, n'}(Z,T) = Z^3 j_{n, n'}(1,T/Z^2).
\label{eq:scaling}
\end{equation}
Using the upper limit on the [O{\sc iii}]~$\lambda$4363~\AA\ line intensity (see Table\,\ref{tab:int})
and $E(B-V)=0.4$ mag, one finds $T\la10\,500$~K. Assuming two times lower line intensity gives
$T\la8900$\,K. We further adopt $T=10\,000$\,K.

Because of the small spatial extent of the He\,{\sc ii}-emitting knot it is likely that radiation emitted 
through recombination into the ground state escapes without being re-absorbed, i.e. case\,A recombination 
can be adopted. For the He\,{\sc ii}\,6560/4686 line ratio we obtain from Martin (1988) that
$j_{\mathrm{HeII}\,6560}/j_{\mathrm{HeII}\,4686} = 0.136$ for case\,A recombination. For case\,B we obtain 
a very similar value. This means that, within the given error margins, the observed
(H$\alpha$+He\,{\sc ii})/He\,{\sc ii}\,4686 line ratio of $0.14\pm0.03$ (see Table\,\ref{tab:int}) can be 
explained by the He\,{\sc ii} emission alone.

For the ratio of H\,$\alpha$/He\,{\sc ii}\,4686 in the He\,{\sc ii}-emitting region we obtain from 
Eq.\,(\ref{eq:scaling}) that $j_{\mathrm{H}\alpha}/j_{\mathrm{HeII}\,4686} = 
[j_{3,2}(1,T)/8 j_{4,3}(1,T/4)](n_{\mathrm{p}}/n_{\mathrm{He^{++}}})$. Assuming that the 
He\,{\sc ii}-emitting zone is fully ionized, we obtain
\begin{equation}
  \label{HAA}
  j_{{\rm H}\,\alpha}/j_{\mathrm{HeII}\,4686} = 0.162(n_\mathrm{H}/n_\mathrm{He})  
\end{equation}
for case\,A recombination, and $0.232 (n_\mathrm{H}/n_\mathrm{He})$ for case\,B. Case\,A thus provides a 
safe lower limit for the H\,$\alpha$ emission. Based on our previous estimate for the He\,{\sc ii} 
emission, the residual contribution of hydrogen to the H$\alpha$ feature is 
$j_{{\rm H}\,\alpha}/j_{\mathrm{HeII}\,4686} < 0.034$. Using Eq.\,(\ref{HAA}), this translates into an 
upper limit of $n_\mathrm{H}/n_\mathrm{He} < 0.2$, meaning that the knot is (almost) free of hydrogen.

\section{More knots around WR\,72}
\label{sec:knots}

\begin{figure}
\begin{center}
\includegraphics[width=8cm,clip=0]{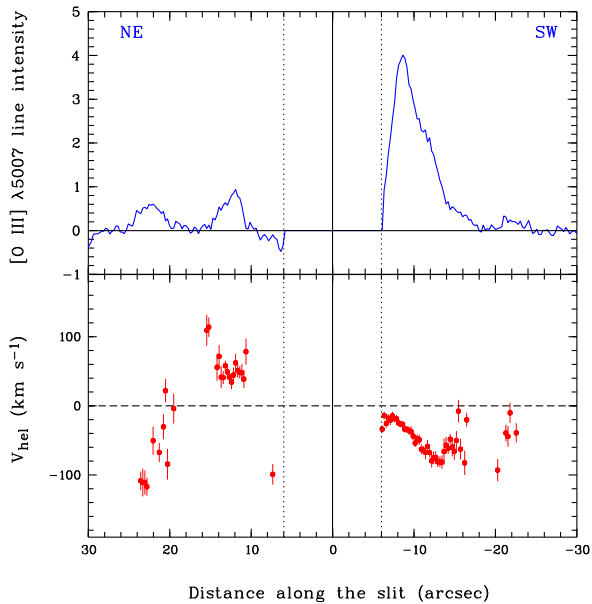}
\end{center}
\caption{[O{\sc iii}]~$\lambda$5007~\AA \, line intensity (upper panel) and the heliocentric radial velocity 
(bottom panel) profiles along the slit. NE--SW direction of the slit is shown. The solid vertical 
line corresponds to the position of WR\,72, while the dashed vertical lines (at $\pm6$ arcsec from the solid 
one) mark the area where the radial velocity was not measured because of the effect of WR\,72.
}
\label{fig:vel}
\end{figure}

\begin{figure*}
\centering
	\includegraphics[width=18cm]{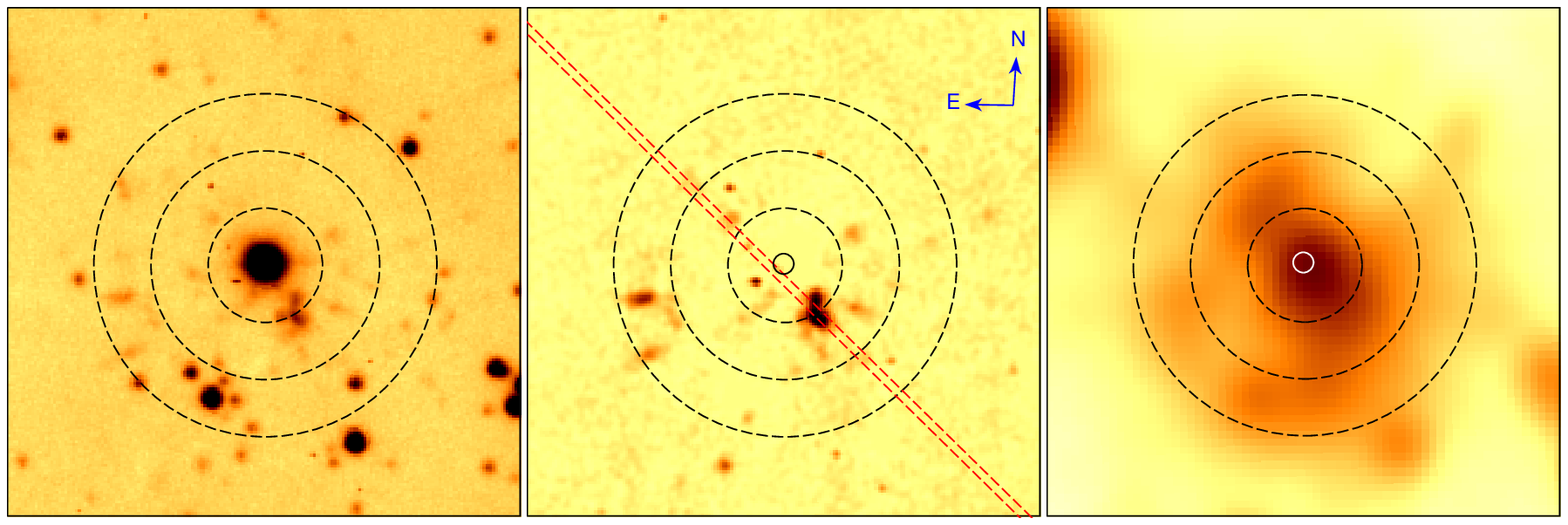}
	\caption{From left to right: SALT [O{\sc iii}]~$\lambda$5007~\AA \, and continuum-subtracted images of 
	(H-poor) knots around WR\,72, and {\it WISE} 4.6\,$\micron$ image of the central shell (with 
	position of WR\,72 indicated by a circle). The orientation and the scale of the images are the same. 
	Concentric, dashed circles of angular radius of 10, 20 and 30 arcsec are overplotted 
	on the images to make their comparison with each other and Fig.\,\ref{fig:vel} more convenient. 
	The location of the 1.5 arcsec wide RSS slit is shown by a (red) dashed rectangle. At the distance 
	to WR\,72 of 1.42 kpc, 30 arcsec correspond to $\approx0.2$ pc.
	}
	\label{fig:salt}
\end{figure*}

Figure\,\ref{fig:vel} plots the [O{\sc iii}]~$\lambda$5007~\AA \, line intensity and the heliocentric 
radial velocity, $v_{\rm hel}$, distributions along the slit. The radial velocity was not measured in the 
area between $-$6 arcsec and +6 arcsec because of the effect of WR\,72. The intensity distribution peaks at 
the position of the already known knot to the southwest of WR\,72 and shows three additional peaks of lower 
intensity, one on the same side from WR\,72 (at $\approx-22$ arcsec), and two other on the opposite side 
(at $\approx+12$ arcsec and +22 arcsec). As we will see below, these peaks correspond to three knots of 
optical emission of low surface brightness, which were intersected by the slit by chance (see 
Fig.\,\ref{fig:salt}). Besides these peaks, we did not detect any signatures of nebular emission within 
the boundaries of the IR halo nor along the whole extent of the 8 arcmin long slit. 

Figure\,\ref{fig:vel} also shows that $v_{\rm hel}$ changes along the slit, suggesting that the 
[O\,{\sc iii}]-emitting material is moving relative to WR\,72. The almost circular shape of 
the IR halo around WR\,72 indicates that the space velocity of the star is much lower than the expansion 
velocity of the halo, meaning that the ram pressure of the interstellar medium did not affect the expansion 
of the stellar wind. Adopting the typical expansion velocity of PNe of $20-30 \, \kms$ (Gesicki \& 
Zijlstra 2000), this condition would imply that the systemic velocity of WR\,72 is close to zero. 
In this case, the radial velocity of the [O\,{\sc iii}]-emitting material with respect to the star is  
given by\footnote{To derive this relation, we used the solar Galactocentric distance 
and the circular Galactic rotation velocity from Reid et al. (2009), and the solar peculiar motion 
from Sch\"onrich, Binney \& Dehnen (2010).}: $v_r =v_{\rm hel}+22.5 \, \kms$. Using this relation and 
Fig.\,\ref{fig:vel}, we speculate that the knot located at +12 arcsec could be an obscured redshifted 
counterpart to the bright knot, just as it was suggested by Pollacco et al. (1992) for the H-poor ejecta 
in the centre of the PN A58 (V605\,Aql). From Fig.\,\ref{fig:vel} and the above relation it would also 
follow that the radial velocity of the [O\,{\sc iii}]-emitting material increases with the distance from 
the central star (cf. Chu et al. 1997; Fang et al. 2014), which is expectable since the stellar wind 
sweeps away and accelerates the material ablated and photoevaporated from the knots (Borkowski et al. 
1995; Chu et al. 1997).

The existence of other knots around WR\,72 is clearly illustrated by Fig.\,\ref{fig:salt}, which shows 
the [O\,{\sc iii}]~$\lambda$5007~\AA \, and continuum-subtracted images of a 1.5 arcmin$\times$1.5 arcmin field 
centred on WR\,72. One can see that besides the bright knot to the southwest from WR\,72 (now resolved into 
two bright blobs) there is a number of faint knots scattered around the star. Some of them are elongated in 
the radial direction like in the fan-like systems of H-poor knots in the central regions of A30 and A78. 
Although the chemical abundances in the newly detected knots have yet to be determined, they apparently
are H-poor as well.

A comparison with the {\it WISE} 4.6\,$\micron$ image shows that all knots are located within the inner shell 
of the IR nebula, and that most of them are enclosed within its southeast lobe (see Fig.\,\ref{fig:salt}).
The linear radius of the shell of $\approx0.2$ pc and the typical radial velocity of knots in the H-poor PNe 
of $\sim100 \, \kms$ (e.g. Chu et al. 1997) imply the age of the shell of $\sim1000$ yr. Thus, WR\,72 is 
of comparable age with the H-poor PNe A30 and A78, and is at a more advanced evolutionary stage than A58, which 
went through the born-again event, respectively, 610--950, 605--1140 and $\sim100$ yr ago (Seitter 1987; 
Fang et al. 2014).
  
\section{Discussion}
\label{sec:dis}

The detection of the H-poor knots within the more extended circular nebula around WR\,72 indicates that this 
star belongs to the rare group of H-poor PNe (only a few such objects are know to date). Using the typical 
expansion velocity of PNe of $20-30 \, \kms$ (Gesicki \& Zijlstra 2000), one finds the kinematic age of the 
IR halo around WR\,72 of 40\,000-60\,000 yr. This age and the large radius of the halo ($\approx1.2$ pc) 
are typical of old PNe (e.g. Pierce et al. 2004), meaning that the lack of optical emission from 
the halo could simply be because its surface brightness has fallen below the detectability threshold (for 
comparison, the radii of the optically visible outer H-rich parts of A30, A58 and A78 are, respectively, 
$\approx0.8$, 0.4 and 0.5 pc). 

The origin of H-poor material in PNe and their central stars is usually attributed to a VLTP, which 
occurs after the extinction of the hydrogen burning shell when the star has already entered the white 
dwarf cooling track. The VLTP leads to the mixing and (almost) complete burning of the remaining hydrogen in
the star, and results in a helium and CO-rich surface composition, which can also be nitrogen-enriched 
(Werner \& Herwig 2006, L\"obling et al. 2019). The chemical composition of WR\,72 has been determined by 
Koesterke \& Hamann (1997) and Keller et al. (2014). Both works consistently find high abundances of carbon 
and oxygen. Additionally, Keller et al. (2014) derive a surface mass fraction of nitrogen of 7 per cent 
for WR\,72, discrepant with a value of 0.5 per cent found by Koesterke \& Hamann (1997). In either case, the 
surface abundances of WR\,72 appear to be consistent with a VLTP evolution.

It is also interesting to use the new {\it Gaia} distance of WR\,72 for constraining its bolometric 
luminosity. For this purpose, we rescale the results of Koesterke \& Hamann (1997) and of Keller et al. 
(2014), who derived the distance to WR\,72 after adopting the bolometric luminosity of this
star. This exercise leads to a luminosity of WR\,72 of 8300 and $22\,000 \, \lsun$, respectively. While 
the larger of the two values is about two times higher than the luminosities of the known [WO]-type CSPNs 
(Gesicki et al. 2006), both are well in the range of luminosities predicted by post-AGB stellar models 
(Bl\"ocker 1995; Miller Bertolami 2016). A high luminosity for WR\,72 could mean that its mass is high 
compared to other [WC]-CSPNe, or that the masses and luminosities of [WC]-CSPNe have been underestimated 
in previous works.

While the VLTP scenario is consistent with the properties of WR\,72, stars with similar properties, with 
C and O rich surfaces, and with hydrogen-deficient circumstellar nebulae, may also arise from the merging 
of two white dwarfs (Longland et al. 2011; Jeffery, Karakas \& Saio 2011; Schwab, Quataert \& Kasen
2016; Schwab 2019). In view of the interpretation of the [WO]-type star IPHAS\,J005311.21+673002.1 as a 
white dwarf merger product (Gvaramadze et al. 2019a) and the potentially high mass of WR\,72 (see above),
it remains possible that also WR\,72 is the result of a white dwarf merger.

Follow-up deeper and higher resolution spectroscopy and imaging of the knots around WR\,72 are needed
to determine their abundances and to check whether their spatial distribution and kinematics are axially 
symmetric. The detection of axial symmetry can be seen as an indication that WR\,72 is a binary system.
On the other hand, during the born-again event, the luminosity of the post-VLTP star could exceed the 
Eddington limit, and thus even slow stellar rotation may break the spherical symmetry of the H-poor ejecta 
(Langer 1997).

Finally, WR\,72 might be a promising target for the Atacama Pathfinder Experiment (APEX) and the current 
X-ray observatories because the bright knot to the southwest of WR\,72 might have a cool molecular core 
(like in A58; Tafoya et al. 2017), while the inter-knot material could be heated to X-ray temperatures by 
the current fast ($2200 \, \kms$; Keller et al. 2014) stellar wind (as takes place in A30 and A78; 
Chu et al. 1997; Guerrero et al. 2012; Toal\'a et al. 2015).

\section{Acknowledgements}
This work is based on observations obtained with the Southern African Large Telescope (SALT), programm 
2019-1-MLT-002. V.V.G. acknowledges support from the Russian Science Foundation under grant 19-12-00383.
A.Y.K. acknowledges support from from the National Research Foundation (NRF) of South Africa. G.G. 
acknowledges financial support from Deutsche Forschunsgemeinschaft (DFG) under grant GR 1717/5-1. This research 
has made use of the NASA/IPAC Infrared Science Archive, which is operated by the Jet Propulsion Laboratory, 
California Institute of Technology, under contract with the National Aeronautics and Space Administration,
the SIMBAD database and the VizieR catalogue access tool, both operated at CDS, Strasbourg, France, and data 
from the European Space Agency (ESA) mission {\it Gaia} (https://www.cosmos.esa.int/gaia), processed by the 
{\it Gaia} Data Processing and Analysis Consortium (DPAC, https://www.cosmos.esa.int/web/gaia/dpac/consortium). 
Funding for the DPAC has been provided by national institutions, in particular the institutions participating in 
the Gaia Multilateral Agreement.

\end{document}